\documentclass[preprint,a4paper]{revtex4-1}

\usepackage[dvipdfmx]{graphicx}
\graphicspath{{./}}

\begin{document}

\title{A solution for secure use of Kibana and Elasticsearch in multi-user environment}

\author{Wataru Takase}
\affiliation{High Energy Accelerator Research Organization (KEK)}
\author{Tomoaki Nakamura}
\affiliation{High Energy Accelerator Research Organization (KEK)}
\author{Yoshiyuki Watase}
\affiliation{High Energy Accelerator Research Organization (KEK)}
\author{Takashi Sasaki}
\affiliation{High Energy Accelerator Research Organization (KEK)}

\begin{abstract}
Monitoring is indispensable to check status, activities, or resource usage of IT services. A combination of Kibana and Elasticsearch is used for monitoring in many places such as KEK, CC-IN2P3, CERN, and also non-HEP communities. Kibana provides a web interface for rich visualization, and Elasticsearch is a scalable distributed search engine. However, these tools do not support authentication and authorization features by default. There is no problem in the case of single-user environment. On the other hand, in the case of single Kibana and Elasticsearch services shared among many users, any user who can access Kibana can retrieve other's information from Elasticsearch. In multi-user environment, in order to protect own data from others or share part of data among a group, fine-grained access control is necessary.

The CERN cloud service group had provided cloud utilization dashboard to each user by Elasticsearch and Kibana. They had deployed a homemade Elasticsearch plugin to restrict data access based on a user authenticated by the CERN Single Sign On system. It enabled each user to have a separated Kibana dashboard for cloud usage, and the user could not access to other's one. Based on the solution, we propose an alternative one which enables user/group based Elasticsearch access control and Kibana objects separation. It is more flexible and can be applied to not only the cloud service but also the other various situations. We confirmed our solution works fine in CC-IN2P3. Moreover, a pre-production platform for CC-IN2P3 has been under construction. 

We will describe our solution for the secure use of Kibana and Elasticsearch including integration of Kerberos authentication, development of a Kibana plugin which allows Kibana objects to be separated based on user/group, and contribution to Search Guard which is an Elasticsearch plugin enabling user/group based access control. We will also describe the effect on performance from using Search Guard.
\end{abstract}

\maketitle

\section{Introduction}
For reliable operation of IT services, a regular monitoring of system status, user activities, and resource usage is essential.
Elasticsearch\cite{elasticsearch} and Kibana\cite{kibana} are open-source monitoring tools developed by Elastic.
Elasticsearch is a distributed, RESTful, horizontally scalable full-text search and analytics engine based on Apache Lucene.
Kibana provides a variety of ways to visualize Elasticsearch data on a web interface.
On the interface, a user can define visualizations of data in Elasticsearch and can create dashboards through arranging and resizing the visualizations.
These visualizations and dashboards are called Kibana objects and stored to an Elasticsearch {\it index} as well as monitoring data.

Elasticsearch and Kibana are used not only in High Energy Physics (HEP) community, such as KEK, CERN, CC-IN2P3 but also in the other fields, such as Facebook, GitHub, Stack Exchange, and so on.
Although these tools provide a useful monitoring platform and are used in many places, they do not support authentication and authorization features by default.
This means any user can access all data in Elasticsearch.
However, in the case that data itself or Kibana objects should be shared with limited users or personalized in some extent, fine-gained access control is necessary.

In this paper, we provide a solution for secure use of Kibana and Elasticsearch in multi-user environment.
This paper starts with an introduction of a solution of the CERN cloud group in Section 2. Section 3 describes our solution consisting of four steps. Section 4 describes performance degradation of secured Elasticsearch. Related work is shown in Section 5, and finally in Section 6, we provide a summary of the work.

\section{Solution of the CERN cloud group}
The CERN private cloud\cite{Bell:scalingthecern} has been in production since 2013.
The cloud is based on OpenStack and configured to be integrated with the CERN network and authentication services.
The cloud service enables a flexible provisioning of computing resources by utilizing virtual machines on demand for the CERN IT services, the batch service, the LHC experimental groups, and personal users.
There are more than two thousands of registered users, and personal and group shared tenants on the cloud.

\begin{figure}[h]
\centering
\includegraphics[width=140mm]{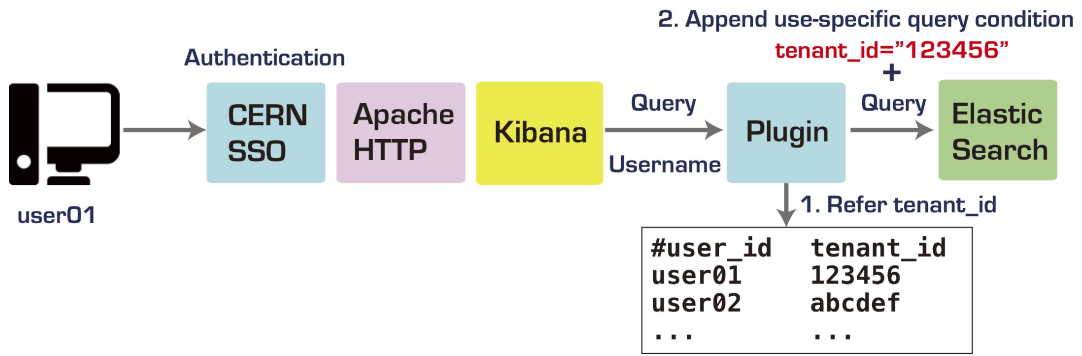}
\caption{Overview of the CERN's solution.}
\label{fig:cern_solution_overview}
\end{figure}

They had provided cloud utilization dashboards to each user by Elasticsearch and Kibana.
On the dashboard, a user could check a usage history of personal tenant such as a number of active virtual machines, used CPU cores, RAMs, storage spaces.
However, a user could also access others' dashboards which contain private information. Moreover, anyone could modify and delete all data in Elasticsearch by default.
To solve the situation, they developed an Elasticsearch plugin to protect user's dashboard from others.
By the plugin, each user could access only information of personal tenants belonging to.
Figure \ref{fig:cern_solution_overview} shows an overview of their solution.
A web server runs in front of Kibana and Elasticsearch to authenticate a user by the CERN Single Sign On system.
After the authentication, the server passes user's request to Kibana.
The developed plugin intercepts a query from Kibana.
Then it appends user-specific query condition based on username and cloud tenant ID that user belongs.
For example, there is a user named {\it user01} who belongs to {\it tenant01} whose tenant ID is {\it 123456}.
If Kibana sends a query requested by user01, the plugin appends {\it tenant\_id=123456} query condition.
Finally, Elasticsearch receives the modified query and returns only results matched the query.

\section{Proposed solution}
Although the CERN's solution enables access control based on user, all the data in Elasticsearch is required to have tenant ID information for the filtering.
Furthermore, the solution is only for their cloud service, and it is difficult to adapt to the other use cases.

Based on their work, we propose a more flexible solution.
It enables fine-grained access control on Elasticserch and Kibana objects separation based on user and groups.
Table \ref{tab:compare_solutions} compares the CERN's and our solutions.
Ours integrates Kerberos 5 for authentication.
Search Guard plugin for Elasticsearch security provides more flexible, various ways of access control than the plugin developed by the CERN cloud group.
In addition, a developed Kibana plugin enables each user or group to have a tenant so that Kibana objects are saved to separated location from others.

\begin{table}
\renewcommand{\arraystretch}{1.3}
\caption{Comparison between the CERN's and our solutions.}
\label{tab:compare_solutions}
\centering
\begin{tabular}{c|c|c}
\hline
& CERN's solution & Our solution\\
\hline
\hline
Authentication & Shibboleth & Kerberos\\
\hline
Elasticsearch plugin & Homemade & Search Guard\\
\hline
\shortstack{\\Elasticsearch\\access control} & \shortstack{\\ \\User based;\\document level} & \shortstack{\\User/group based; index, type,\\document,operation level}\\
\hline
\shortstack{\\Kibana object\\separation} & \shortstack{Per user\\ \vspace{1mm} } & \shortstack{\\ \\Per user/group enabled by\\a developed Kibana plugin}\\
\hline
\end{tabular}
\end{table}

\begin{figure}[h]
\centering
\includegraphics[width=140mm]{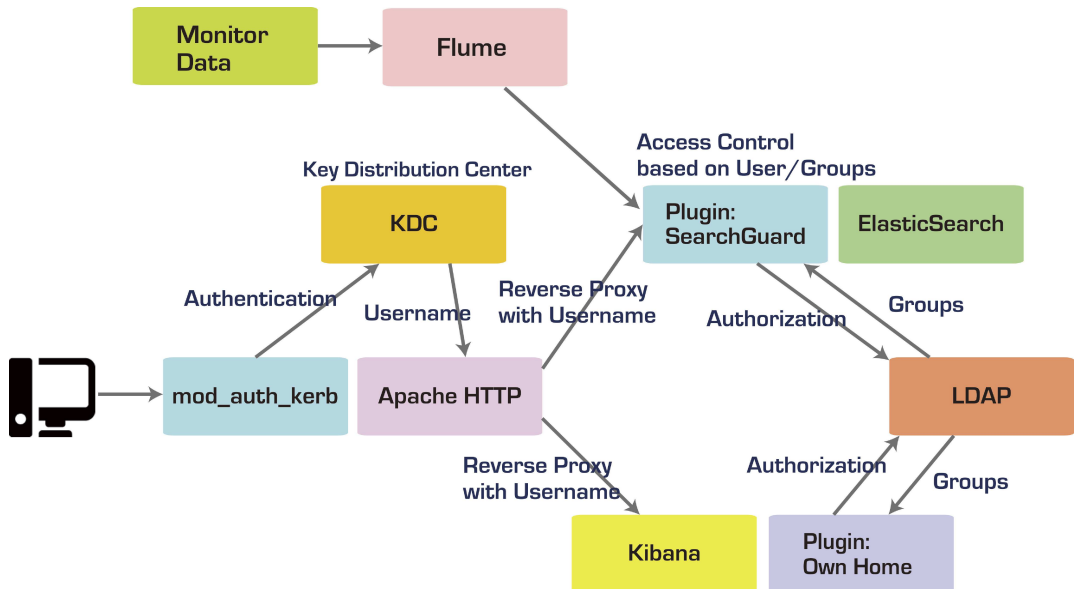}
\caption{Overview of our solution.}
\label{fig:overview_of_solution}
\end{figure}

Figure \ref{fig:overview_of_solution} shows an overview of our solution.
Kibana and Elasticsearch run behind of a web server.
Therefore these services only allow access from authenticated user.
The developed Kibana plugin generates available Kibana tenant list based on user and LDAP groups.
A user can switch a tenant on a web interface depending on the situation.
All requests going to Elasticsearch are intercepted by Search Guard plugin.
The plugin enables user and LDAP groups based access control on Elasticsearch.
Apache Flume is a tool which collects system logs and metrics from servers and stores them to Elasticsearch.
Our Flume patch enables to push data to Search Guard-enabled Elasticsearch over SSL/TLS connection.

Our solution can replace the CERN's solution and moreover be adaptable for the other various use cases by securing Kibana and Elasticsearch.
The monitoring group in CC-IN2P3, Lyon, France, have tried to centralize their Elasticsearch and have investigated the way of access control on Elasticsearch.
We collaborated with them under TYL-FJPPL (Toshiko Yuasa Laboratory France-Japan Particle Physics Laboratory) Comp\_03  project, and a prototype of our solution worked well in CC-IN2P3 in February 2016.
Also, a production platform has been under construction.

The following sub-sections describe each part of our solution.

\subsection{Integration of Kerberos authentication}
For the first step, we set up a web server using Apache HTTP in front of Kibana and Elasticsearch.
The server is an only entry point to access the both services.
We configured to use Kerberos 5 authentication on the server.
The other authentication method is adaptable such as Basic, LDAP, Shibboleth by changing authentication module of Apache HTTP.
After authentication, the web server set an authenticated username to HTTP request header and then passes a request to the behind service so that Kibana or Elasticsearch recognizes who requested.

\subsection{Development of a Kibana plugin for multi-teancy}
In the case of single Kibana instance shared among many users and groups, all Kibana objects (searches, visualizations, dashboards) are stored to the same Elasticsearch index which is called {\it Kibana index}.
This means any user can access, modify, and also delete all objects in the index.
In multi-user environment, to protect own objects from others or share part of objects among a group, Kibana index separation is one of the solutions.

One idea is preparing Kibana instances as many as users.
Each user has a dedicated Kibana instance configured to have a different Kibana index from the other instances.
However, this idea does not scale.
In the case that there are thousands of Kibana users, the same amount of Kibana instances is necessary.

The developed plugin named Own Home\cite{ownhome} adds multi-tenancy feature to single Kibana instance.
The plugin enables a user to have own personal Kibana index so that objects the user created are stored to separate location from others.
Furthermore, a group shared Kibana index can be provided.
A user can switch Kibana index depending on his/her use cases.
Available Kibana index list is generated based on username, LDAP groups, and file based definition.

\begin{figure}[h]
\centering
\includegraphics[width=140mm]{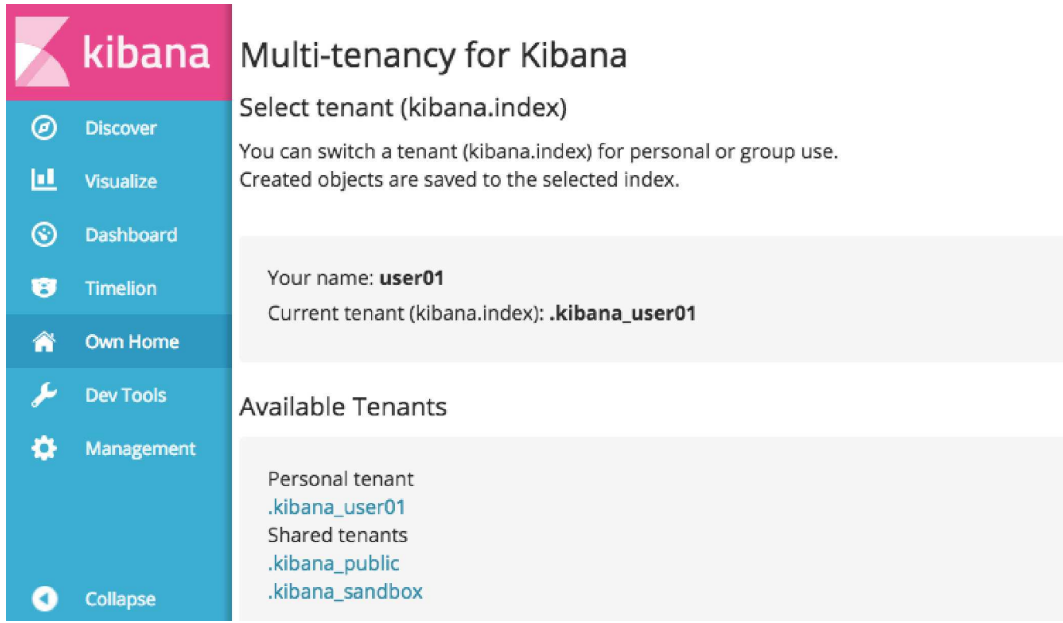}
\caption{Screen-shot of Own Home.}
\label{fig:own_home}
\end{figure}

Figure \ref{fig:own_home} shows a screen-shot of the plugin.
A current selected Kibana index is stored in HTTP session.
When a user switches Kibana index by clicking one of available Kibana indices, the plugin updates session to newly selected one.

\begin{figure}[h]
\centering
\includegraphics[width=140mm]{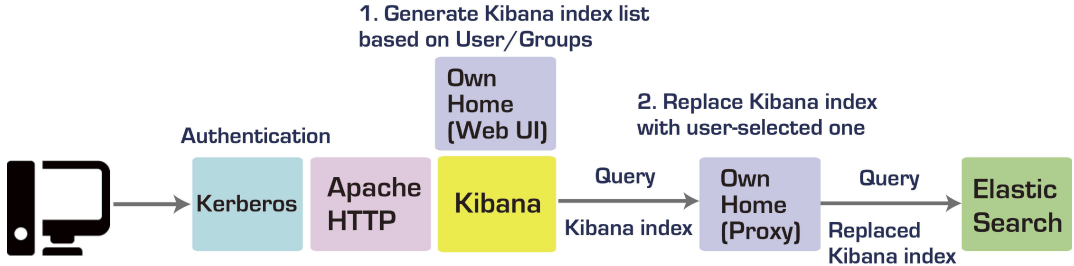}
\caption{Workflow of Kibana index replacement.}
\label{fig:own_home_workflow}
\end{figure}

Figure \ref{fig:own_home_workflow} shows how the plugin works.
The plugin launches a proxy server on the same host of Kibana instance at the time of initialization.
At first, a user selects a Kibana index from available index list on the web interface.
Although Kibana always requests Elasticsearch to store Kibana objects into the single Kibana index, the proxy server intercepts the requests, then replaces the Kibana index with user's selected one which comes from HTTP session.

\subsection{Configuration and contribution to Search Guard}
The developed Kibana plugin just separates Kibana index and stores Kibana objects to a different location based on user/group.
For the next step, it is necessary to set index level access control on Elasticsearch.
For example, only {\it user01} can access {\it user01}'s Kibana index, or users belonging to {\it group01} can access {\it group01}'s index.

Search Guard\cite{searchguard} is an Elasticsearch plugin for security developed by Floragunn.
The software is open-source under Apache license.
It provides flexible REST/Transport layer access control based on user/group, indices, types, documents, or cluster operations.
Moreover, Elasticsearch node-to-node connections are encrypted for security.
Search Guard supports multiple authentication and authorization backends such as Basic, Kerberos, proxy-based authentication, and LDAP authorization.
In the case of our solution, authentication has been done at the web server in front of Elasticsearch.
Hence, proxy-based authentication is selected as an authentication method.
Also, LDAP authorization is configured for access control based on user and LDAP groups.
An Elasticsearch admin can define access control list and push it to a highly secured Elasticsearch index.
When a user accesses to data in Elasticsearch, Search Guard checks the user's permission by the list.

In addition, we have contributed to Search Guard community by proposing some code patches for more flexible configuration.
All of the patches have been merged into the upstream code.
The following describes one of the patches.
In the case that each user has own Kibana index and each index allows access only from the owner, an admin has to define permissions for every user.
Furthermore, whenever a new user is registered, the admin has to add permission and update the access control list.
Our patch enables to use username variable in access control list, so access is controlled dynamically accessed user basis.

\subsection{Development of an Apache Flume patch for SSL/TLS connection}
Apache Flume\cite{flume} is one of the tools to push monitoring data to Elasticsearch.
It enables to collect, aggregate, and move a large amount of data from many different sources to a centralized data store.
Flume provides several kinds of {\it sinks} which put collected data to an external repository like HDFS.
{\it Flume Elasticsearch sink} is one of them enabling to push collected data to Elasticsearch via plain text connection.
However, the sink request is refused by Search Guard because Search Guard encrypts Elasticsearch connections and does not allow non-SSL/TLS requests.
Our Flume patch enables to support SSL/TLS connection, and we confirmed the patched Flume could push data to Search Guard-enabled Elasticsearch.

\section{Evaluation of Search Guard-enabled Elasticsearch performance}
The previous section described securing Elasticsearch, and this section looks into its impact on performance.
We measured Elasticsearch performance degradation caused by Search Guard using Rally\cite{rally}.
Rally is a benchmarking tool for Elasticsearch developed by Elastic.
It measures indexing throughput, query latency, aggregation latency, stats latency, and so on.
Rally provides a few default benchmark scenarios, and also a user can define customized one.

In this paper, we used one of the default scenarios named {\it geonames}.
In the scenario, Rally downloads geographical dataset (see Table \ref{tab:data_structure}) from a data source, then indexes the documents, total 2.8 GB, using eight client threads against a target Elasticsearch cluster.
\begin{table}
\renewcommand{\arraystretch}{1.3}
\caption{Part of data structure of the geographical dataset used in Rally.}
\label{tab:data_structure}
\centering
\begin{tabular}{c|c|c}
\hline
Field & Type & Description\\
\hline
\hline
name & string & Name of geographical point\\
\hline
country\_code & string & ISO-3166 2-letter country code\\
\hline
population & long & Population\\
\hline
longitude & double & Longitude\\
\hline
latitude & double & Latitude\\
\hline
\end{tabular}
\end{table}

\begin{figure}[h]
\centering
\includegraphics[width=140mm]{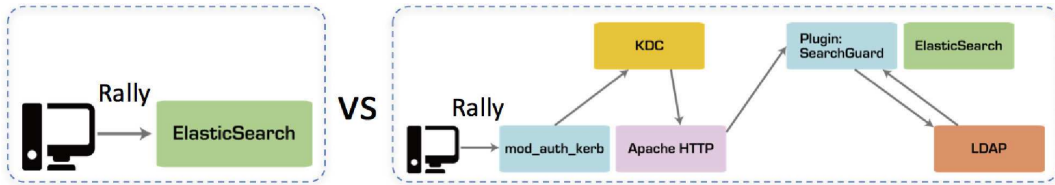}
\caption{Elasticsearch performance comparison by Rally.}
\label{fig:rally_test_scenario}
\end{figure}

We prepared two physical machines of AMD Opteron 6212 2.6 GHz 8 cores and 8 GB of RAMs with CentOS 7 and set up an Elasticsearch cluster on top of them.
One of the machines also serves as a web server, Kerberos Key Distribution Center, LDAP server in the case of Search Guard-enabled environment.
We used Rally 0.3.1, and Elasticsearch and Search Guard versions are 2.3.4.
Then normal and Search Guard-enabled Elasticsearch performances were compared.
For normal Elasticsearch measurement, Rally accesses Elasticsearch REST API directory.
On the other hand, for Search Guard-enabled environment, Rally accesses a front end web server(see Figure \ref{fig:rally_test_scenario}).

\begin{figure}[htbp]
\begin{tabular}{cc}
\begin{minipage}{0.5\hsize}
\begin{center}
\includegraphics[width=6cm]{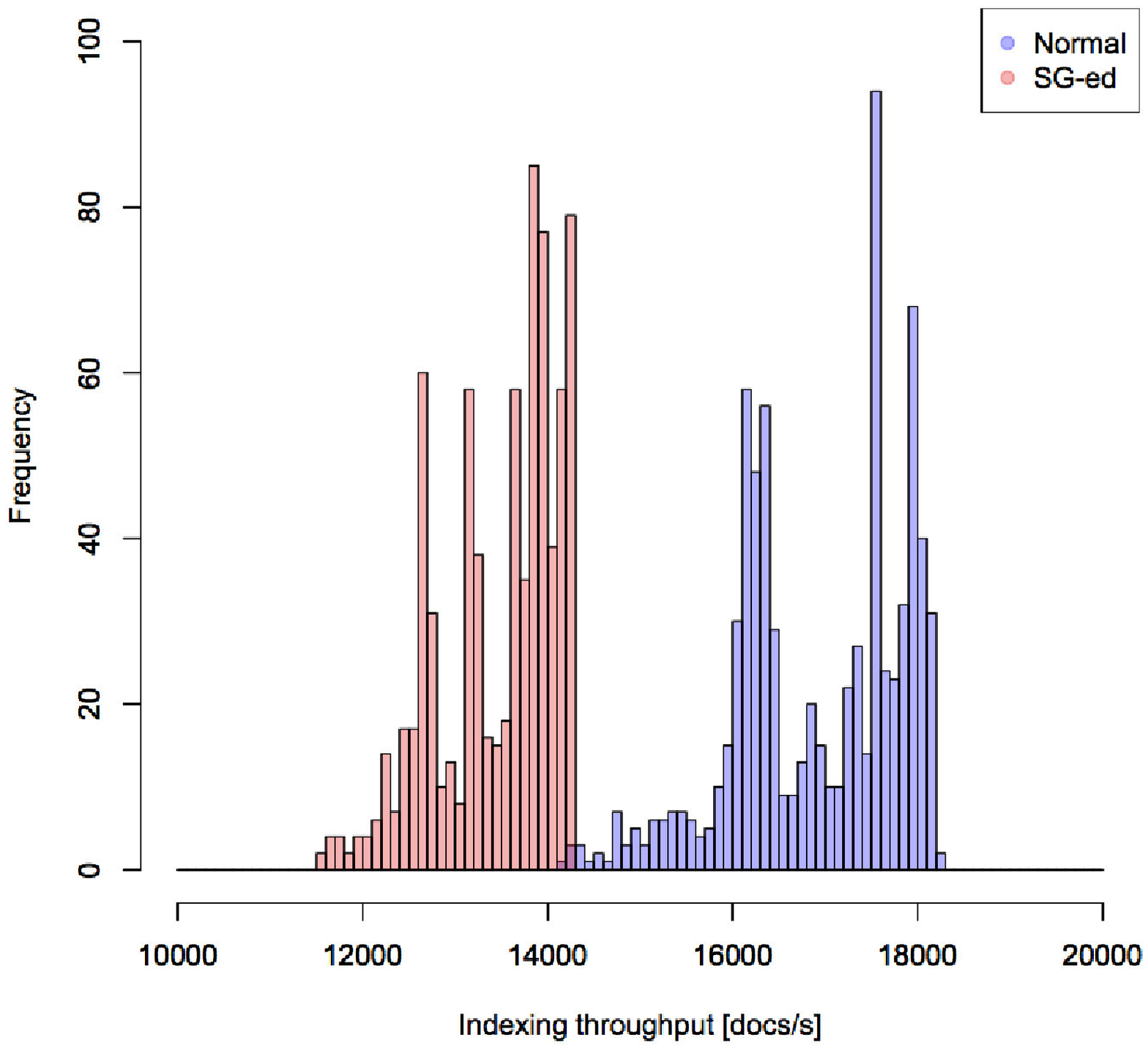}
\caption{Distributions of indexing throughput}
\label{fig:indexing_throughput}
\end{center}
\end{minipage}
\begin{minipage}{0.5\hsize}
\begin{center}
\includegraphics[width=6cm]{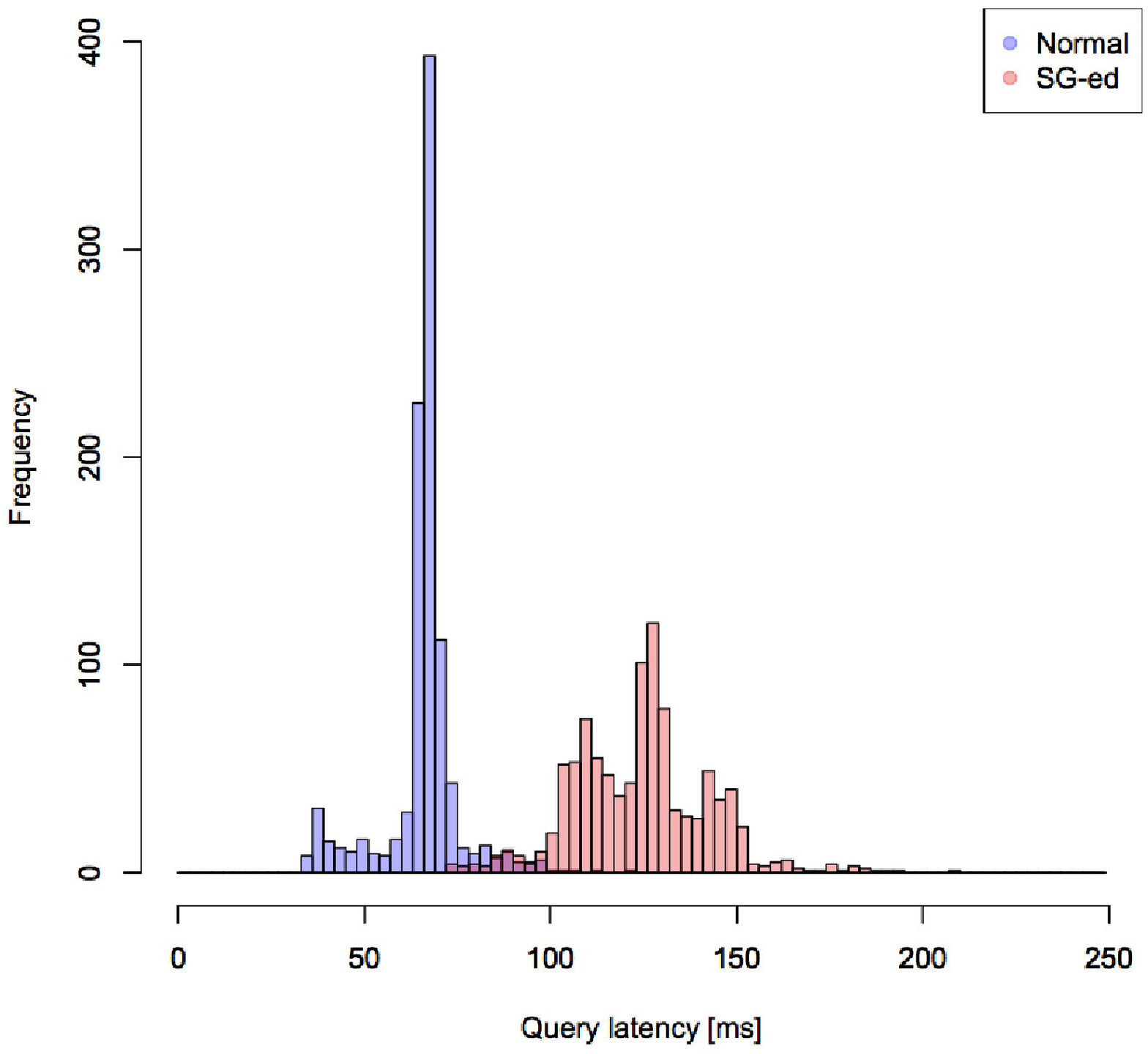}
\caption{Distributions of default query latency}
\label{fig:query_latency_default}
\end{center}
\end{minipage}
\end{tabular}
\end{figure}

\begin{figure}[htbp]
\begin{tabular}{cc}
\begin{minipage}{0.5\hsize}
\begin{center}
\includegraphics[width=6cm]{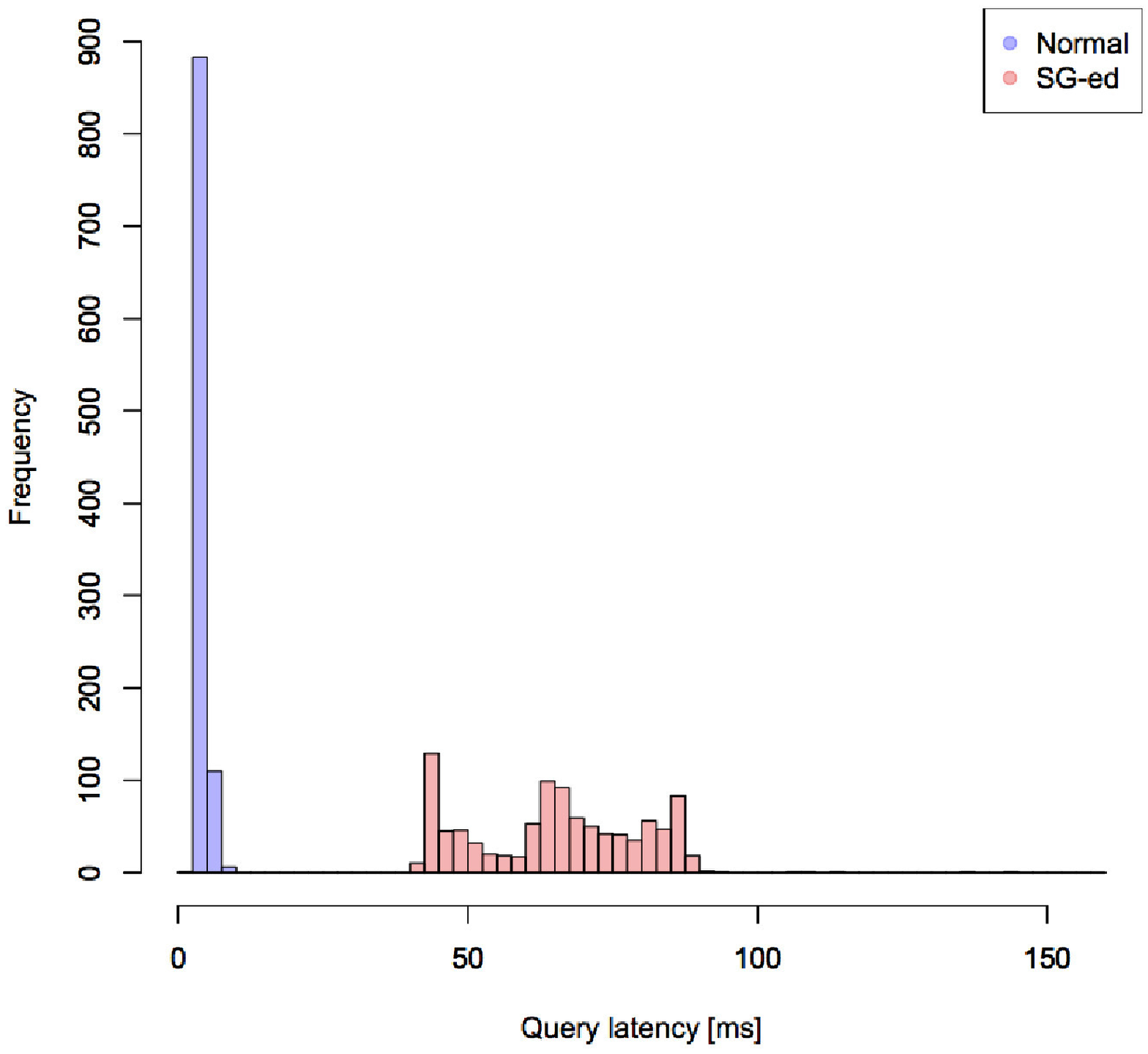}
\caption{Distributions of term query latency}
\label{fig:query_latency_term}
\end{center}
\end{minipage}
\begin{minipage}{0.5\hsize}
\begin{center}
\includegraphics[width=6cm]{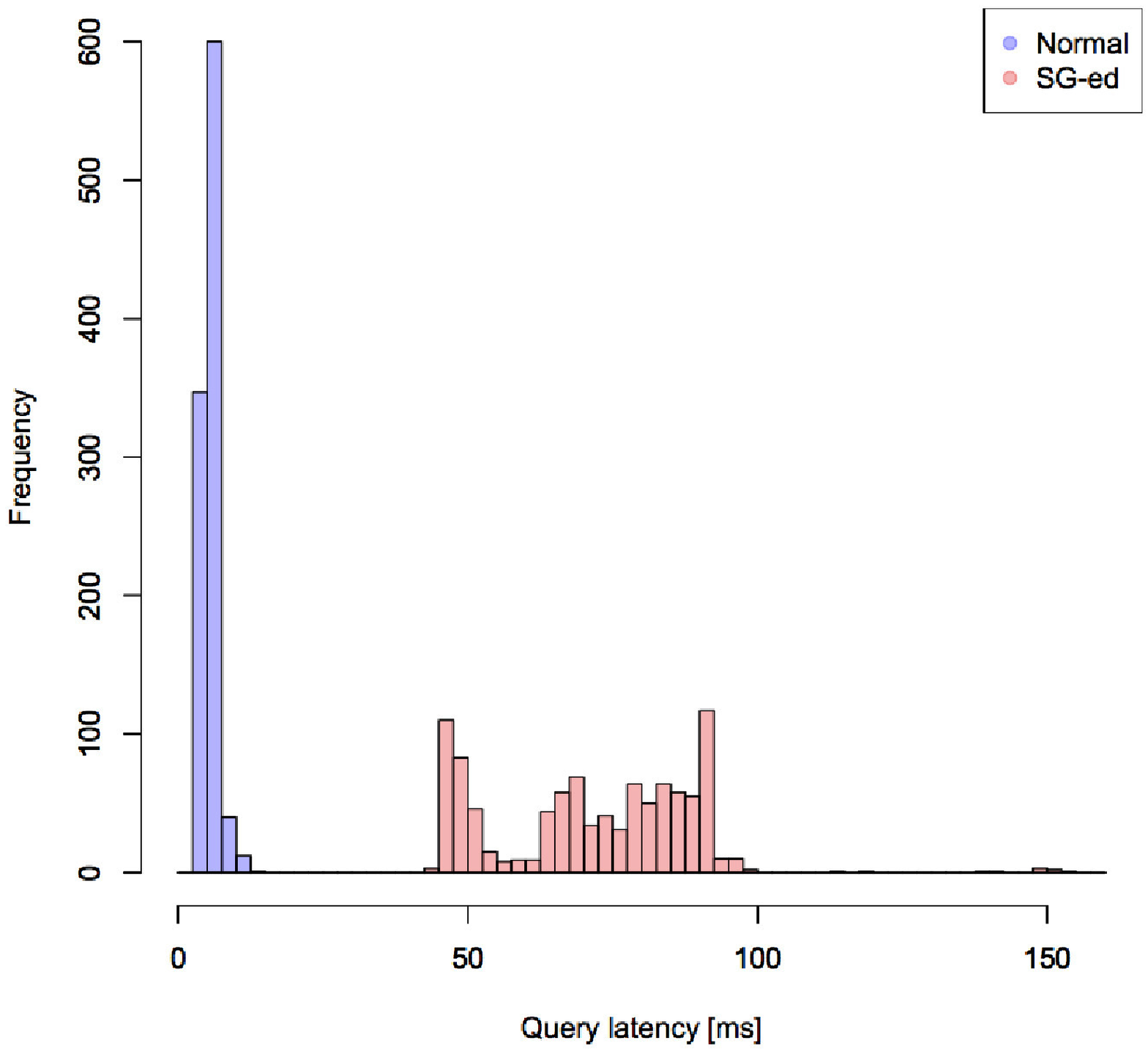}
\caption{Distributions of phrase query latency}
\label{fig:query_latency_phrase}
\end{center}
\end{minipage}
\end{tabular}
\end{figure}

\begin{figure}[htbp]
\begin{tabular}{cc}
\begin{minipage}{0.5\hsize}
\begin{center}
\includegraphics[width=6cm]{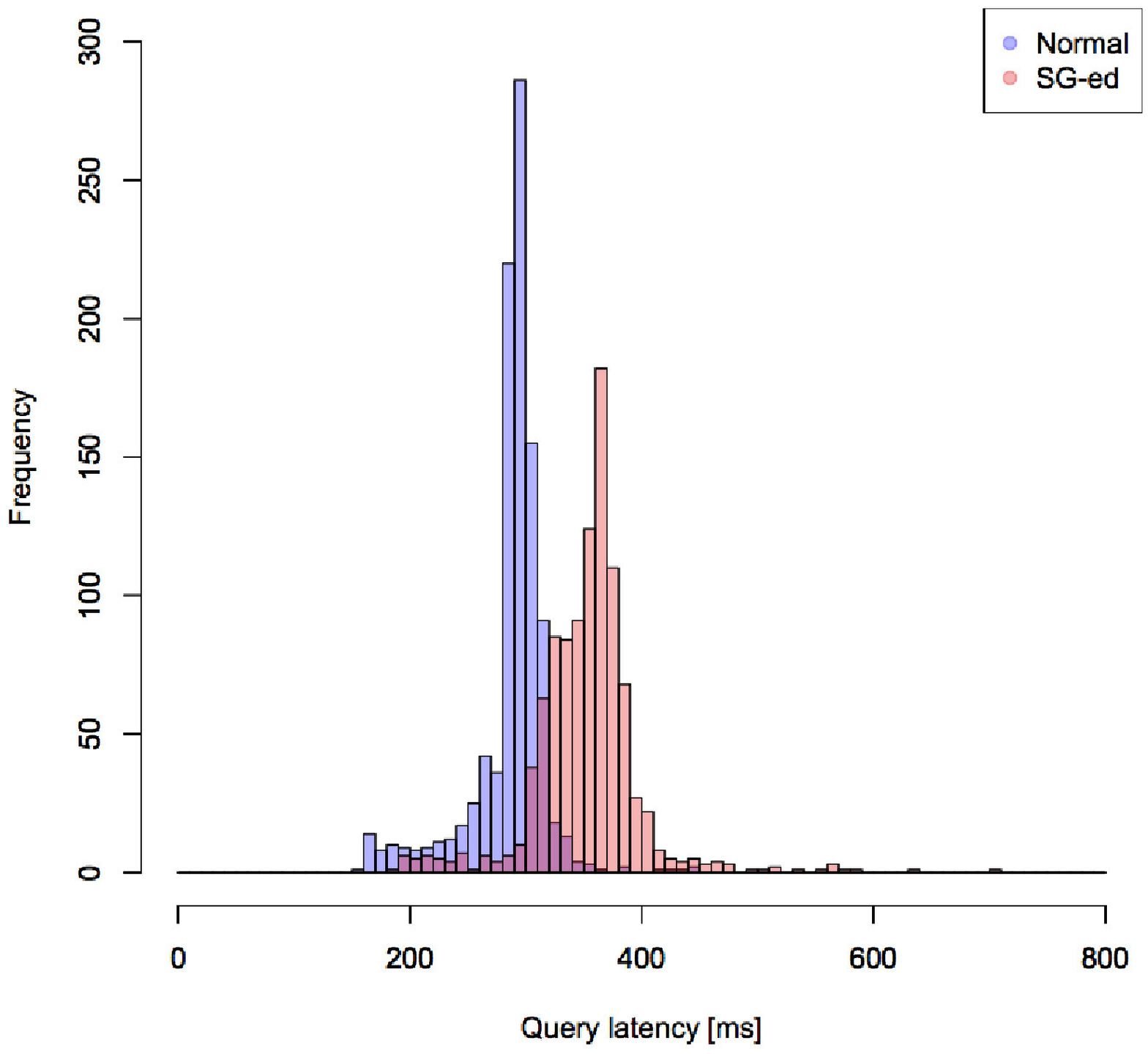}
\caption{Distributions of aggregation without caching query latency}
\label{fig:query_latency_agg_uncached}
\end{center}
\end{minipage}
\begin{minipage}{0.5\hsize}
\begin{center}
\includegraphics[width=6cm]{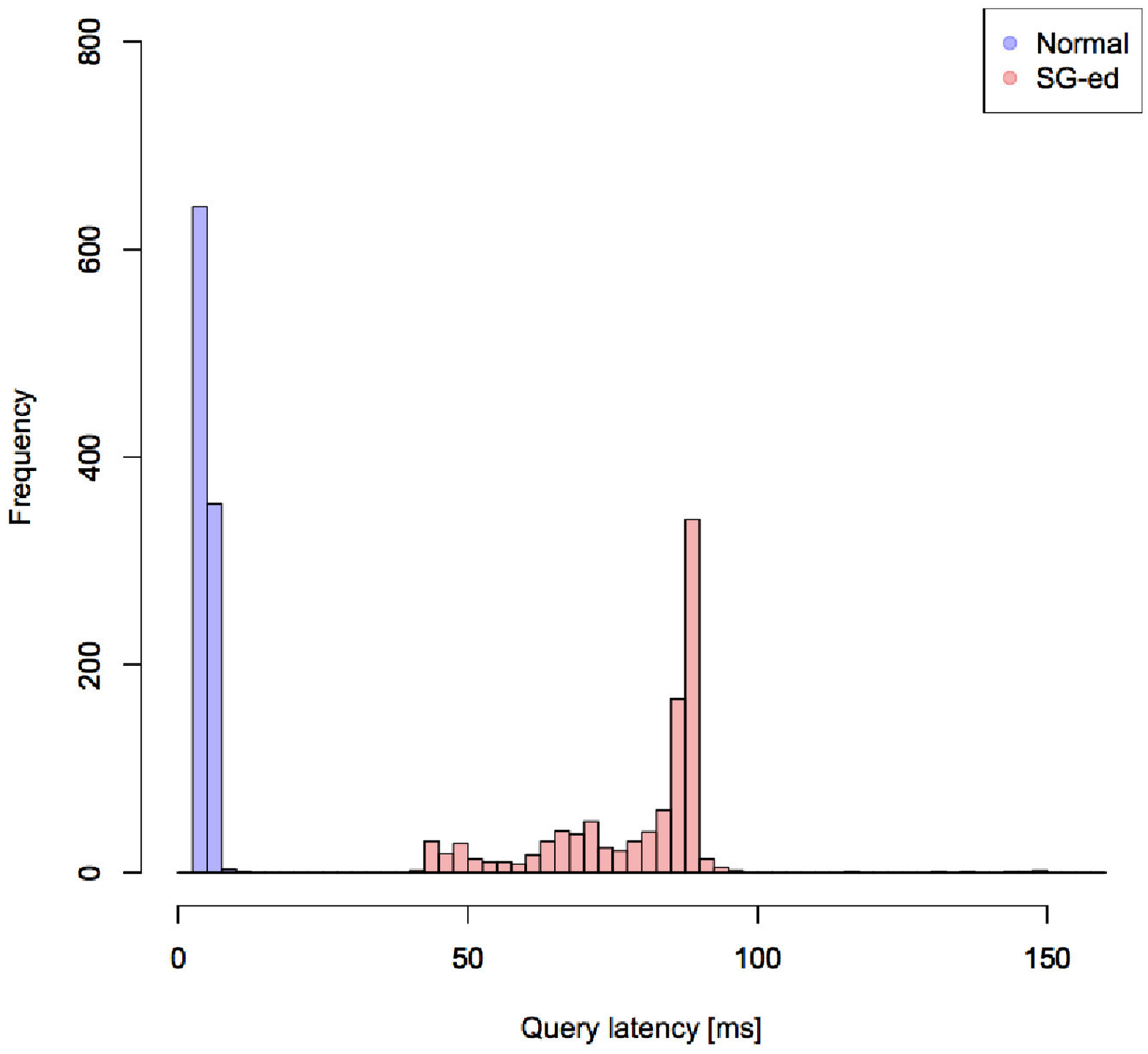}
\caption{Distributions of aggregation with caching query latency}
\label{fig:query_latency_agg_cached}
\end{center}
\end{minipage}
\end{tabular}
\end{figure}

\begin{figure}[htbp]
\begin{tabular}{cc}
\begin{minipage}{0.5\hsize}
\begin{center}
\includegraphics[width=6cm]{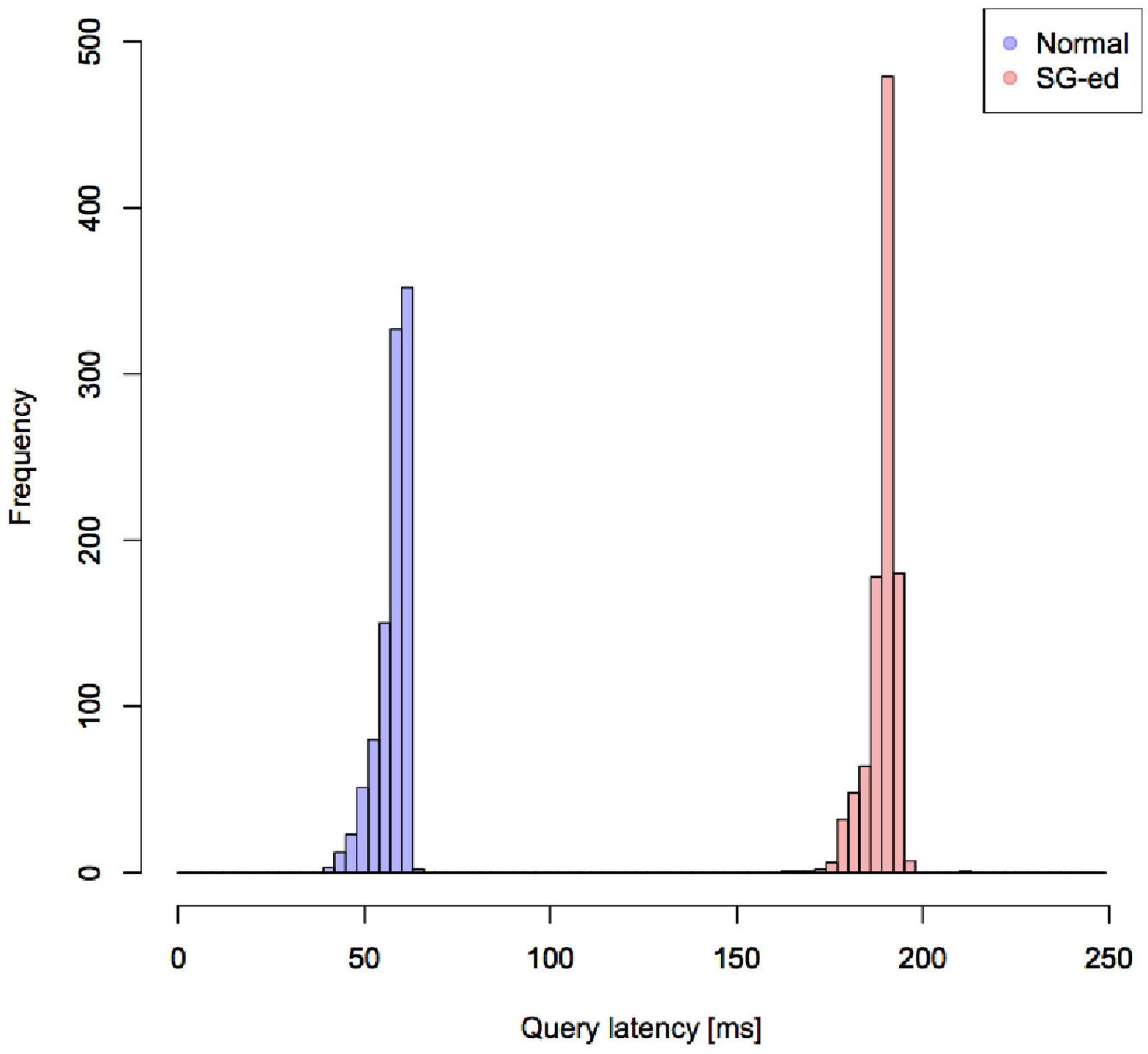}
\caption{Distributions of scroll query latency}
\label{fig:query_latency_scroll}
\end{center}
\end{minipage}
\begin{minipage}{0.5\hsize}
\begin{center}
\includegraphics[width=6cm]{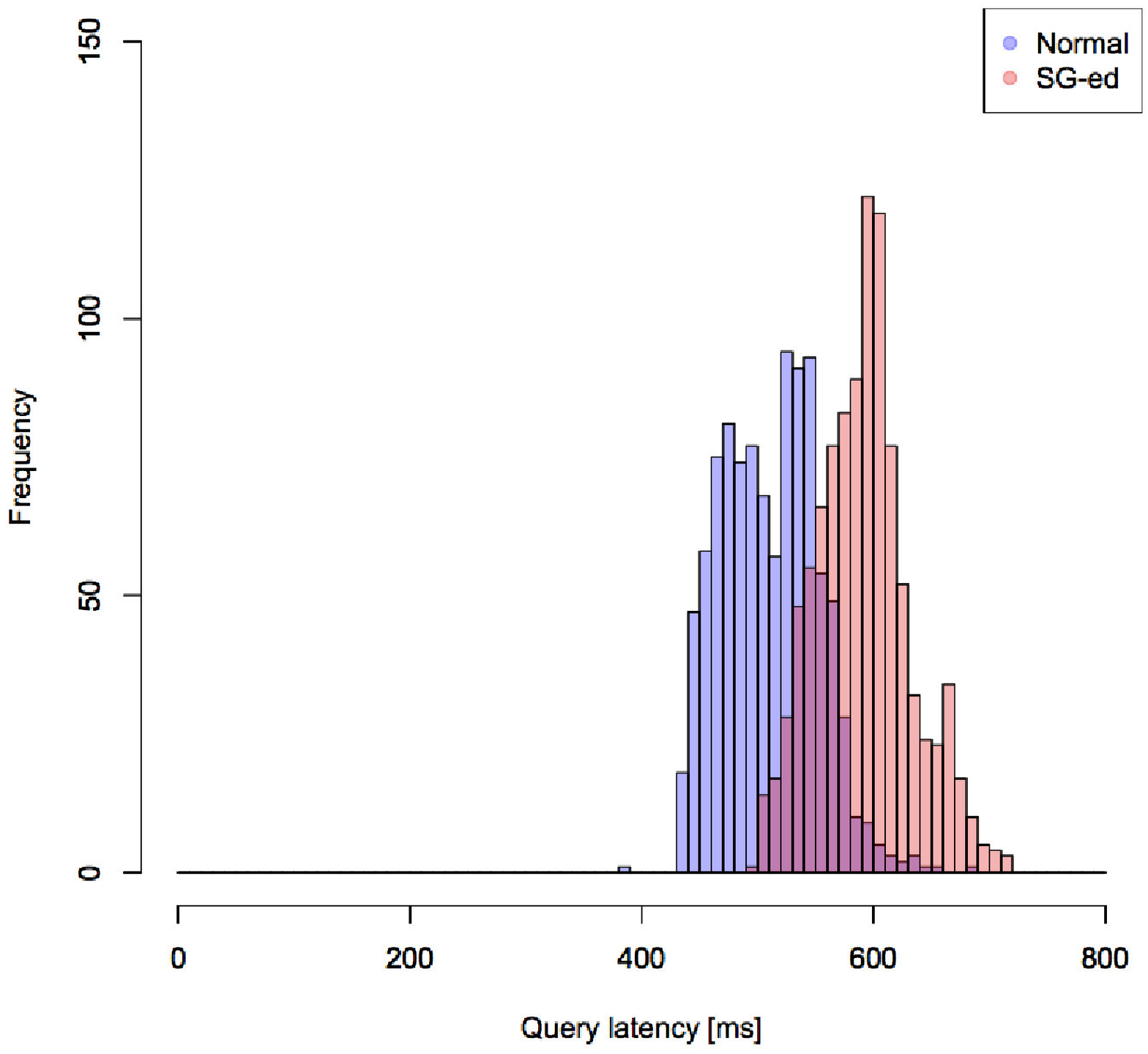}
\caption{Distributions of expression query latency}
\label{fig:query_latency_expression}
\end{center}
\end{minipage}
\end{tabular}
\end{figure}

We measured indexing throughput and several kinds of query latencies.
Figure \ref{fig:indexing_throughput}-\ref{fig:query_latency_expression} show the throughput or latency distributions of the measured benchmark result.
The blue and red correspond to the normal and Search Guard-enabled Elasticsearch results respectively.

Figure \ref{fig:indexing_throughput} shows distributions of indexing throughput.
Rally repeated to push 5000 documents per request and stored 8.6 million documents in total.
The test measured the time from storing a document to making it searchable.
Figure \ref{fig:query_latency_default}-\ref{fig:query_latency_expression} show query latency results and 1000 queries were executed in each of the measurements.
Distributions of seven kinds of query latency results are shown respectively.
Table \ref{tab:query_description} shows descriptions of each of the queries.
\begin{table}
\renewcommand{\arraystretch}{1.3}
\caption{Descriptions of queries.}
\label{tab:query_description}
\centering
\begin{tabular}{c|c}
\hline
Query & Description\\
\hline
\hline
Default query & Searches all documents.\\
\hline
Term query & Searches documents whose {\it country\_code} equal to {\it AT}.\\
\hline
Phrase query & Searches documents whose {\it name} contain {\it Sankt Georgen}.\\
\hline
\shortstack{\\ \\Aggregation query\\with/without caching} & \shortstack{\\Sums up {\it population} grouped by\\{\it country\_code} with/without caching.}\\
\hline
\shortstack{Scroll query\\ \vspace{0.4mm}} & \shortstack{\\ \\Searches all documents, and fetches\\the results 1000 docs per page 25 times.}\\
\hline
\shortstack{Expression query\\ \vspace{0.4mm}} & \shortstack{\\ \\Searches all documents, and evaluates each doc by\\calculation using {\it population}, {\it elevation}, and {\it latitude}.}\\
\hline
\end{tabular}
\end{table}

Comparisons of medians between normal and Search Guard-enabled results are shown in Table \ref{tab:median}.
We observed 20\% of performance degradation of index throughput in the case of using Search Guard.
Moreover, regarding the query latencies, there are certain differences depending on queries.
Overhead of each query except the scroll is estimated as around 60 to 80 ms by the medians in Table \ref{tab:median}, and overhead of the scroll query is about 130 ms.
These degradations are caused by Kerberos authentication, reverse proxy, LDAP lookup, and Search Guard operations.

Normal Elasticsearch cannot be shared among users who have different purposes because of lack of access control feature.
Secured Elasticsearch enables to provide a centralized service which can cover various use cases instead of launching Elasticsearch instance for each case.
It leads to reduce maintenance cost, simplify system structure, and use limited compute resource effectively.
However, it causes some performance deterioration.
This paper provides a criterion of the degradation in secured Elasticsearch environment.
In the case that a user requires higher indexing throughput and lower latency than Search Guard-enabled environment, setting up his/her dedicated Elasticsearch cluster is one of the solutions instead of using shared Elasticsearch among others.

\begin{table}
\renewcommand{\arraystretch}{1.3}
\caption{Comparisons of medians between nomal and Search Guard-enabled Elasticsearch.}
\label{tab:median}
\centering
\begin{tabular}{c|c|c|c}
\hline
& \shortstack{\\Normal\\Elasticsearch} & \shortstack{\\Search Guard\\enabled Elasticsearch} & \shortstack{\\Perfomance\\degradation}\\
\hline
\hline
Indexing throughput [docs/s] & 17076 & 13659 & 3417 (20\%)\\
\hline
Default query latency [ms] & 67.0 & 125.0 & 58.0\\
\hline
Term query latency [ms] & 3.8 & 65.5 & 61.7\\
\hline
Phrase query latency [ms] & 5.4 & 73.4 & 68.0\\
\hline
\shortstack{\\ \\Aggregation without caching\\query latency [ms]} & \shortstack{\\292.1\\ \vspace{0.4mm}} & \shortstack{\\357.4\\ \vspace{0.4mm}} & \shortstack{\\65.3\\ \vspace{0.4mm}}\\
\hline
\shortstack{\\ \\Aggregation with caching\\query latency [ms]} & \shortstack{\\4.5\\ \vspace{0.4mm}} & \shortstack{\\86.2\\ \vspace{0.4mm}} & \shortstack{\\81.7\\ \vspace{0.4mm}}\\
\hline
Scroll query latency [ms] & 58.9 & 190.3 & 131.4\\
\hline
Expression query latency [ms] & 510.3 & 592.0 & 81.7\\
\hline
\end{tabular}
\end{table}

\section{Related work}
Bagnasco et al. \cite{Bagnasco:monitoringofiaas} set up a web server in front of Kibana in order to authenticate and authorize a user at the web server level.
However, They did not mention about access control on Elasticsearch.
Furthermore, all Kibana objects are accessible from authenticated users just by setting up a proxy server.

There is another Elasticsearch plugin for security named X-Pack Security (formerly known as Shield)\cite{shield} maintained by Elastic.
It provides IP filtering, authentication, authorization, node encryption, and auditing.
The plugin is not only for Elasticsearch but also well integrated with Kibana.
On Kibana interface, an admin can define users and roles for access control.
The plugin is a commercial product and also 30 days free trial license is available.
It provides access restriction features almost the same as Search Guard. However, Search Guard provides additional features such as OpenSSL support, Kerberos support, HTTP Proxy authentication support, JSON Web token support.
On Kibana side, X-Pack Security does not provide multi-tenancy feature such our developed Kibana plugin supports.
Therefore, in the case of using X-Pack Security, all users still can access others Kibana objects.

\section{Summary}
For secure use of Kibana and Elasticsearch, this paper provides a solution which enables user/group based access restriction and Kibana object separation in multi-user environment.
A web server authenticates a user and passes requests to backends.
A developed Kibana plugin allows a user to switch Kibana index depending on the situation such as personal use or group shared use.
Search Guard enables user/group based access control on Elasticsearch, and we have contributed to Search Guard community for more flexible configuration.
A developed patch for Apache Flume enables SSL/TLS connection so that Flume pushes data to Search Guard-enabled Elasticsearch.
We measured the effect on the performance of Search Guard environment. The result shows that indexing throughput performance is degraded 20\%, the overhead of each query except the scroll is estimated as around 60 to 80 ms, and the scroll query overhead is about 130 ms.
It would be acceptable in an interactive use of monitoring system.

\bibliographystyle{JHEP}
\bibliography{references}

\end{document}